\documentclass[runningheads]{llncs}
\usepackage[T1]{fontenc}
\usepackage{graphicx}
\usepackage{amsmath}
\usepackage{multirow}
\usepackage{booktabs}
\usepackage{hyperref}

\usepackage{tabularx}
\usepackage[style=numeric,sorting=none]{biblatex}
\begin{document}
\title{Exploring Communication Strategies for Collaborative LLM Agents in Mathematical Problem-Solving\thanks{The manuscript has been accepted for presentation at the 26th International Conference on Artificial Intelligence in Education (AIED 2025), to be held from July 22–26, 2025, in Palermo, Italy.}}


\author{Liang Zhang \inst{1,2}\orcidID{0009-0002-0017-2569} \and
Xiaoming Zhai\inst{3,4,5}\orcidID{0000-0003-4519-1931} \and
Jionghao Lin\inst{6}\orcidID{0000-0003-3320-3907} \and Jennifer Kleiman\inst{3,5}\orcidID{0009-0000-3171-1954} \and Diego Zapata-Rivera \inst{7} \orcidID{0000-0002-0620-7622} \and Carol Forsyth \inst{7} \orcidID{0000-0003-4830-5156} \and Yang Jiang \inst{7} \orcidID{0000-0002-2195-5776} \and Xiangen Hu \inst{8} \orcidID{0000-0001-9045-4070} \and  Arthur C. Graesser \inst{10} \orcidID{0000-0003-0345-6866}
}
%

\institute{Institute for Intelligent Systems, University of Memphis, Memphis, TN, USA \and
Department of Electrical and Computer Engineering, University of Memphis, Memphis, TN, USA \\ \email{\{lzhang13\}@memphis.edu} \and AI4STEM Education Center, University of Georgia, Athens, GA, USA \and National GENIUS Center, Athens, GA, USA \and Department of Mathematics, Science, and Social Studies Education, University of Georgia, Athens, GA, USA
\\
\email{\{Xiaoming.Zhai\}@uga.edu} \and Faculty of Education, The University of Hong Kong, Hong Kong, PR China \\ \and Educational Testing Service, Princeton, NJ, USA \and Department of Applied Social Sciences, Hong Kong Polytechnic University, Hong Kong, PR China
\and Department of Psychology, University of Memphis, Memphis, TN, USA
}

\maketitle              
%


\begin{abstract} Large Language Model (LLM) agents are increasingly utilized in AI-aided education to support tutoring and learning. Effective communication strategies among LLM agents improve collaborative problem-solving efficiency and facilitate cost-effective adoption in education. However, little research has systematically evaluated the impact of different communication strategies on agents' problem-solving. Our study examines four communication modes, \textit{teacher-student interaction}, \textit{peer-to-peer collaboration}, \textit{reciprocal peer teaching}, and \textit{critical debate}, in a dual-agent, chat-based mathematical problem-solving environment using the OpenAI GPT-4o model. Evaluated on the MATH dataset, our results show that dual-agent setups outperform single agents, with \textit{peer-to-peer collaboration} achieving the highest accuracy. Dialogue acts like statements, acknowledgment, and hints play a key role in collaborative problem-solving. While multi-agent frameworks enhance computational tasks, effective communication strategies are essential for tackling complex problems in AI education.  

\end{abstract}

\section{Introduction}

Recent advances in Large Language Models (LLMs) have shown significant promise in enhancing educational tasks such as feedback generation, learning guidance, interaction strategies, understanding student behaviors, and fostering tutoring dialogues through answer evaluation and content generation \cite{zhang2024predicting,zhang2025data,zhang2024spl,tan2023does,lee2024applying}. However, most implementations rely on single-agent approaches using prompt engineering for logical reasoning (e.g., chain of thought \cite{wei2022chain}). Emerging research on multi-agent collaboration among LLMs further improves performance through collaborative strategies and holds promise for expanding the effectiveness of AI in education \cite{yu2024mooc,yue2024mathvc}. In multi-agent settings, LLM agents leverage complementary reasoning processes to cross-validate ideas, refine their understanding, and ultimately produce more robust solutions than single-agent approaches \cite{latif2024systematic}. Despite these promising developments, systematic exploration of communication strategies among collaborative LLM agents in problem-solving contexts remains limited. 

Mathematical problem-solving presents an ideal testing ground for multi-agent LLM collaboration in education. The domain of mathematics demands precise, step-by-step reasoning and offers clear, binary outcomes, making it well-suited for evaluating how multiple LLM agents can effectively reason, communicate, and learn together. Recent studies on chat-based math problem solving \cite{liang2024mathchat} and conversation programming for mathematical problem solving \cite{keating2024zero} illustrate the potential of multi-agent approaches to address the unique challenges of mathematical reasoning. Systematic analysis of diverse problem-solving strategies in math can provide valuable insights into optimizing agent communication and enhancing overall performance in AI-driven educational systems \cite{zhang2022exploring}. 

\begin{figure*}[ht!]
    \centering
\includegraphics[width=1\textwidth]{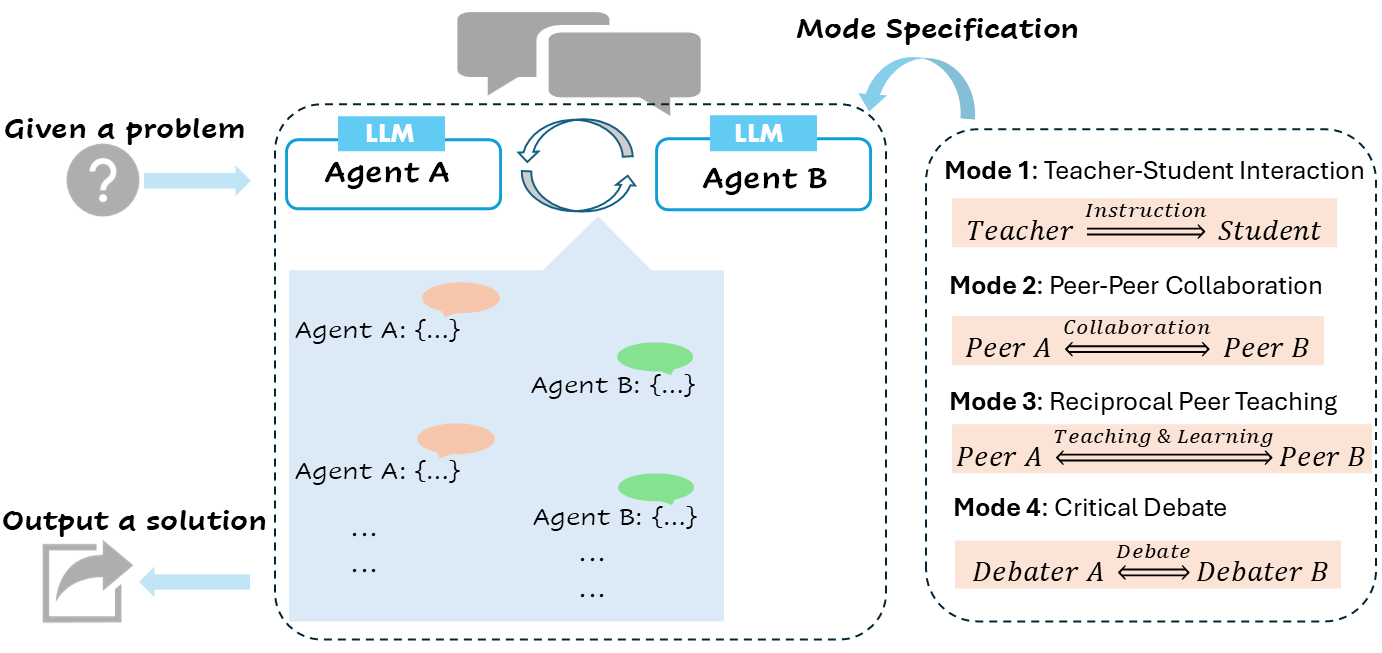}
    \caption{An illustration for different dual LLM agents modes for problem-solving.}
    \label{fig:dual_agent} 
\end{figure*}

In this study, we explore a range of collaborative communication strategies using dual LLM-based agents to solve mathematical problems via chat-based interactions. Drawing inspiration from Vygotsky’s educational theories, particularly the Zone of Proximal Development (ZPD) and the More Knowledgeable Other (MKO) \cite{vygotsky1978mind}, and informed by insights from machine psychology \cite{hagendorff2023machine}, we adopt several interaction modes in our study. These include \textit{teacher-student interactions}, where one agent guides another through direct instruction and the Socratic method; \textit{peer-to-peer collaboration}, in which agents work as equals to exchange ideas and solve problems; \textit{reciprocal peer teaching}, where agents alternate roles to reinforce understanding; and \textit{critical debate}, where agents challenge one another’s solutions to refine their approaches. Fig. \ref{fig:dual_agent} provides further details on these communication strategies. By assigning distinct roles to the agents, we aim to enhance their collaborative problem-solving effectiveness and evaluate the performance of these communication strategies in math problem-solving. Dialogue Act (DA) analysis will be applied to conversations across different strategies to identify unique traits and derive actionable insights. Our investigation is guided by the following \textbf{R}esearch \textbf{Q}uestions: 
\begin{itemize}
    \item \textbf{RQ 1:} Which communication strategies among dual-agent LLM systems most effectively enhance mathematical problem-solving performance?  
    \item \textbf{RQ 2:} How do DA patterns vary across dual-agent communication modes and influence mathematical problem-solving? 
\end{itemize} 

The explorations and findings will guide the development of collaborative multi-agent systems powered by LLMs, paving the way for more adaptive and human-centric AI educational platforms. This research contributes to creating knowledgeable agents that leverage effective collaborative strategies to drive advanced intelligent tutoring systems. Ultimately, these insights promise to bridge fundamental AI research with practical educational applications, thereby enhancing learning outcomes in AI-aided education.

\section{Method}
All primary language model configurations in LLM agent design were based on the OpenAI GPT-4o model \cite{hurst2024gpt}.

\textbf{Dataset.} We used a subset (a total of 700) of the MATH dataset \cite{hendrycksmath2021}, selecting only level-5 problems (the highest difficulty level) \footnote{\url{https://huggingface.co/datasets/hendrycks/competition_math}}.

\textbf{Dual-Agent Protocol Definition.} We define four dual-agent communication protocols, each tailored to a distinct communication mode (see Fig. \ref{fig:dual_agent}). Each protocol involves two LLM agents, \(\mathcal{A}\) and \(\mathcal{B}\), to whom we assign specific personas and provide comprehensive guidance for response generation via prompt engineering. For example, in the \textit{teacher–student interaction} setup, the agent $\mathcal{A}$ adopts the teacher persona with the summarized version of prompt template \(\mathcal{T}_A(Q,\mathcal{H}) = \) ``\textit{You are a supportive math teacher. Present \(Q\) clearly, summarize key points, offer hints, and ask probing questions based on dialogue history \(\mathcal{H}\) without revealing the full solution}'', and the agent $\mathcal{B}$ adopts the student persona with the summarized version of prompt template \(\mathcal{T}_B(Q, \mathcal{H})= \) ``\textit{You are a curious math student. Follow the teacher’s guidance to solve \(Q\) step-by-step, articulate your reasoning, and ask for clarifications as needed.}'' For further details, including comprehensive agent definitions and prompt templates for alternative communication modes, please refer to our GitHub repository:\url{https://github.com/LiangZhang2017/aied_dual_agent}.

\textbf{Baseline.} We established a baseline single-agent protocol in a zero-shot setting. The single agent is defined as a helpful assistant tasked with solving math problems, generating both the solution process and the final answer by sequentially working through each problem. 

\textbf{Accuracy Evaluation.} The final answer produced by both dual-agent and single-agent setups is rigorously compared to the ground truth to assess correctness. Accuracy is defined as the percentage of correctly answered questions, normalized to a 100-point scale for each math subject, communication mode, and run. Average accuracy is computed across different communication modes, with the standard error (SE) calculated from three independent runs to account for observed inconsistencies in the final answers. 

\textbf{Dialogue Act Classification Analysis.} The conversation exhibits in-depth reasoning that reveals the underlying logical structure of mathematical problem solutions. To further examine the nuances of dialogue communication and explore the distinct functionalities of dual agents in collaborative math problem solving, we perform a dialogue act classification analysis. Dialogue acts refer to the functional roles that utterances or responses play in a conversation \cite{lin2022good, lin2023robust, vail2014identifying}. Dialogue acts can include actions like asking questions, providing explanations, and offering feedback. Initially, each agent's response is segmented into chunks, where a chunk may consist of a single sentence, a series of strongly logically related sentences, or a comprehensive multi-step equation process. We then used a pre-trained Bert-based model introduced in \cite{lin2022good}, which has been trained and validated on large-scale tutoring dialogue data (3,629 utterances) collected from math tutoring sessions between human tutors and students. Additionally, we apply a framework based on the principles from \cite{vail2014identifying}, which provides a structured method for categorizing different types of dialogue acts. Below is a brief reference list of DA tags from the scheme in \cite{vail2014identifying}: H (Hint), DIR (Directive), ACK (Acknowledge), RC (Request Confirmation), RF (Request Feedback), PF (Positive Feedback), NF (Negative Feedback), LF (Lukewarm Feedback), Q (Question), A (Answer), and S (Statement). 



\section{Results and Discussion}

\subsection{Performance Accuracy in Four Communication Modes} 

\begin{table}[ht]
    \centering
    \scriptsize
    \caption{Average Accuracy (\%) across Different Modes with SE.}
    \begin{tabular}{lcccccc}        \toprule
        \multirow{2}{*}{\textbf{Model}} & \multirow{2}{*}{\textbf{Single Agent}} & \multicolumn{4}{c}{\textbf{Dual Agents}} \\
        \cmidrule(lr){3-6}
         &  & \textbf{Teacher-Student} & \textbf{Peer-to-Peer} & \textbf{Critical Debate} & \textbf{Reciprocal Peer} \\
        \midrule
        GPT-4o & \(47.43_{3.66}\) & \(52.33_{4.13}\) & \textbf{\(54.10_{3.91}\)} & \(52.76_{4.14}\) & \(51.95_{3.95}\) \\
        
        \bottomrule
    \end{tabular}
\label{tab:total_avg_correctness}
\end{table}

Table \ref{tab:total_avg_correctness} displays the average accuracy for each communication mode, calculated by averaging the performance across all math problems solved within that mode. The results demonstrate that employing dual-agent performance in mathematical problem-solving can yield significant improvements over single-agent settings. Notably, the \textit{peer-to-peer collaboration} mode, where agents work as equals by sharing intermediate results and cross-verifying outputs, achieved the highest average accuracy (54.10\%). Collaborative contribution through reciprocal interactions, as observed in this mode, significantly boost error-checking and reasoning, both of which are crucial for tackling complex mathematical tasks \cite{li-etal-2023-theory, bo2025reflective}. Moreover, the \textit{peer-to-peer collaboration} mode exhibited the lowest standard error (SE = 3.91) among all dual-agent approaches, signifying more robust and consistent performance. The teacher-student interaction (52.33\%), critical debate (52.76\%), and reciprocal peer teaching (51.95\%) modes also yield improvements over the single agent (47.43\%), although their differences are relatively small.

\subsection{Dialogue Act Classification Result}

\begin{figure*}[h!t]
\centering
\includegraphics[width=4.9in]{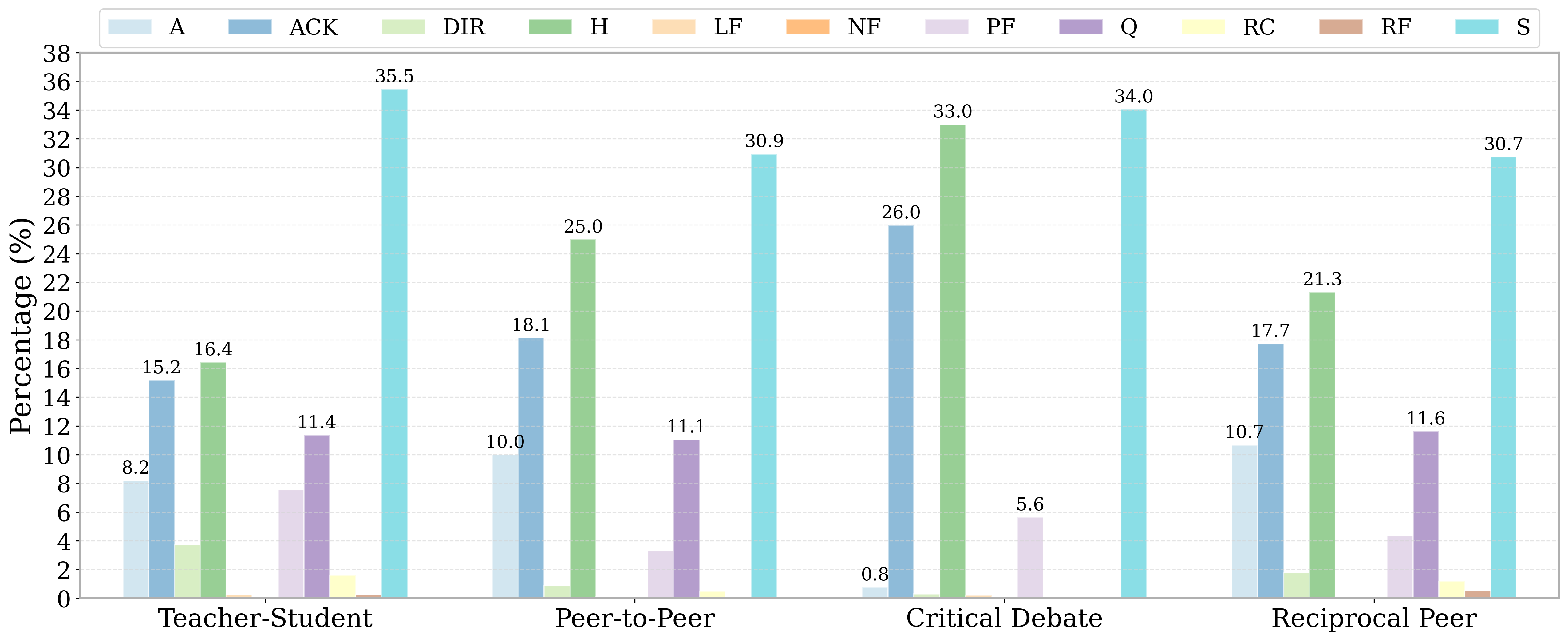}
\caption{DA Classification Results Across Four Modes.}
\label{fig:dialogue_act_percentage_by_mode}
\end{figure*} 

Fig.~\ref{fig:dialogue_act_percentage_by_mode} illustrates the frequency distribution of DAs across four communication modes, exemplified by the algebra subject. Both agents' utterances were merged to capture their collective impact on problem solving, and the top five DA tags in each mode are highlighted in the upper histogram. Notably, statement (S), hints (H) and acknowledge (ACK) emerge as the three most frequent dialogue acts across all modes, significantly contributing to collaborative problem-solving. In particular, statement (S) is the predominant DA across all four modes, with its peak frequency (35.5\%) observed in the teacher-student interaction mode. For example, an agent might state, ``\textit{The algebraic manipulation and the verification steps are clear and logical},'' or ``\textit{However, I would like to explore the solution further to ensure there are no other possible pairs of x and y that satisfy the given conditions.}'' These examples illustrate how agents express their own perspectives and guide the subsequent moves in problem solving. The hints (H) emerge as the most frequent dialogue act in critical debate (33.0\%), where agents actively challenge each other’s solutions. The utterances ``\textit{Let's denote the two whole numbers I pick as x and y.}'' and ``\textit{According to the problem, my friend picks the numbers x-4 and 2y-1.}'' serve as hints that are intended to guide the problem-solving process rather than provide complete solutions. This high prevalence of hints reflects an active effort to explore, test, and improve upon proposed solutions \cite{vanlehn2006behavior}. Acknowledge (ACK), for instance, ``\textit{Thank you, Agent B, for your systematic approach to verifying the solution,}'' or ``\textit{Thank you for the constructive discussion!}'' serve as bridging elements that affirm successful communication between agents. These acknowledgments not only recognize individual contributions but also strengthen the overall collaborative dynamic. Additionally, answer (A) are among the top five most frequently occurring DAs. Overall, the high frequency of these dual-agent-specific dialogue acts underscores their crucial role in driving the collaborative problem-solving process. Using Kendall's correlation \cite{kendall1938new}, we found that DA patterns in peer-to-peer collaboration are highly correlated with teacher-student interaction (0.99) and reciprocal peer teaching (0.96), but less so with critical debate (0.67), suggesting that even subtle variations in these patterns can significantly impact performance.

\section{Conclusion}
In this study, we explored the impact of multi-agent communication strategies on mathematical problem-solving using LLM-based tutoring agents. By systematically comparing four distinct dual-agent communication modes including \textit{teacher-student interaction}, \textit{peer-to-peer collaboration}, \textit{critical debate}, and \textit{reciprocal peer teaching}, we demonstrated that multi-agent frameworks can significantly enhance problem-solving performance relative to single-agent approaches. Notably, the \textit{peer-to-peer collaboration} mode achieved the highest accuracy, underscoring the benefits of equal, collaborative interaction among agents. Our dialogue act analysis revealed that specific communicative behaviors, such as statements (S), acknowledge (ACK), and hints (H), play pivotal roles in the agents' collaborative reasoning process. These findings offer valuable insights into how different interaction patterns contribute to effective problem solving and provide a foundation for designing more adaptive AI tutoring systems. Future work will aim to refine agent communication protocols, integrate advanced prompt optimization techniques, and explore heterogeneous agent configurations to further enhance performance. Overall, our work demonstrates that structured, multi-agent communication not only boosts accuracy in mathematical problem-solving but also enriches the underlying dialogue dynamics, paving the way for more robust and human-centric AI educational platforms. 

\section{Acknowledgments}
The research reported here was partially supported by the Institute of Education Sciences, U.S. Department of Education, through Grant R305C240010 to University of Georgia Research Foundation. The opinions expressed are those of the authors and do not represent views of the Institute or the U.S. Department of Education.

\printbibliography


\end{document}